\documentclass[aps,pra,superscriptaddress,twocolumn,amsmath,amssymb,floatfix]{revtex4-1}

\usepackage{times}

\usepackage[utf8x]{inputenc}
\usepackage[english]{babel}
\usepackage[T1]{fontenc}
\usepackage{lmodern}
\usepackage{amssymb}
\usepackage{bbold}
\usepackage{amsmath}
\usepackage{amsthm}
\usepackage[pdftex]{graphicx}
\usepackage{epstopdf}

 \usepackage{bm}
 \usepackage{relsize,dsfont,mathrsfs,empheq,verbatim,upgreek}

\newcommand{\Tr}{{\mathrm{Tr}}}

\newcommand{\trn}[1]{|| #1 ||}

\begin{document}

\title{Mixing-induced quantum non-Markovianity and information flow}


\author{Heinz-Peter Breuer}
\affiliation{Physikalisches Institut, Universit\"at Freiburg, 
Hermann-Herder-Stra{\ss}e 3, D-79104 Freiburg, Germany}

\author{Giulio Amato}
\affiliation{Physikalisches Institut, Universit\"at Freiburg, 
Hermann-Herder-Stra{\ss}e 3, D-79104 Freiburg, Germany}
\affiliation{Dipartimento di Fisica, Universit\`a degli Studi di Milano,
Via Celoria 16, I-20133 Milan, Italy}

\author{Bassano Vacchini}
\affiliation{Dipartimento di Fisica, Universit\`a degli Studi di Milano,
Via Celoria 16, I-20133 Milan, Italy}
\affiliation{INFN, Sezione di Milano, Via Celoria 16, I-20133 Milan, Italy}

\begin{abstract}
Mixing dynamical maps describing open quantum systems can lead from Markovian 
to non-Markovian processes. Being surprising and counter-intuitive, this result has been 
used as argument against characterization of non-Markovianity in terms of information
exchange. Here, we demonstrate that, quite the contrary, mixing can be understood 
in a natural way which is fully consistent with existing theories of memory effects. 
In particular, we show how mixing-induced non-Markovianity can be interpreted in terms 
of the distinguishability of quantum states, system-environment correlations and the 
information flow between system and environment.
\end{abstract}
 

\date{\today}

\maketitle

\textit{Introduction.}  In quantum as well as in classical mechanics
the isolation of the system of interest is never perfectly achievable. 
The effect of external noise or of the interaction with uncontrolled environmental 
degrees of freedom makes the dynamics stochastic. In quantum mechanics the
environmental influence appears as an additional layer of
stochasticity, on top of the inherently probabilistic description of
any quantum experiment, and cannot be generally described by means of
classical stochastic processes. Quantum processes, which can be taken
as the description of the evolution of an open quantum system dynamics
\cite{Breuer2007}, are described by time dependent collections of
completely positive trace preserving (CPT) maps, called quantum
dynamical maps. The characterization of quantum processes in view of
the relationship of these maps at different times, in analogy to the
correlations of a classical process at different times,
which allow to introduce the very definition of Markovian process,
is an important and difficult issue due to the special role of measurement
in quantum mechanics. 

In recent times a lot of work has been devoted to the study of quantum 
non-Markovianity (see e.g. \cite{Breuer2012a,Rivas2014a,Breuer2016a,Devega2017a}). 
In particular, a notion of memory for quantum processes has been
introduced which can be physically interpreted in terms of the flow of
information between the open system and its environment
\cite{Breuer2009b,Wissman2015a}. The information flow is defined by means of
the behavior in time of the distinguishability of two open system states and
non-Markovianity is characterized by a non-monotonic time evolution of the
distinguishability. Experimental control and measurements of non-Markovian
quantum dynamics and of the closely connected impact of initial system-environment 
correlations have been reported for photonic systems
\cite{Liu2011a,Li2011a,Tang2012a,Liu2013a,Cialdi2014a,Bernardes2015,Tang2015},
nuclear magnetic resonance \cite{Bernardes2016}, and trapped ion systems
\cite{Gessner2014a,Wittemer2017}.

However, mixing of quantum
dynamical maps leads to new time evolutions, whose Markovianity properties
can be related in a quite counter-intuitive way to the Markovianity of
the original maps
\cite{Vacchini2012a,Chruscinski2013a,Chruscinski2015a,Wudarski2015a,Kropf2016a,
Wudarski2016a}. In particular one can consider random mixtures of
unitary evolutions showing up memory effects, so that objections have
been raised about the validity of the interpretation of
non-Markovianity in terms of information flow
\cite{Pernice2012a,Megier2017a}. On the contrary, we show in this contribution by means 
of a microscopic description that non-Markovianity arising by mixing is naturally understood in 
terms of information flow. This implies in particular that indeed the collection of
quantum dynamical maps giving the reduced system dynamics allows to properly describe
memory effects.

\textit{Mixing quantum processes.}
We consider two quantum processes given by one-parameter families of quantum dynamical
maps $\Phi^{(1)}_t$ and $\Phi^{(2)}_t$ with $t\geq 0$. 
A natural way to construct a new map is to consider the convex linear combination
\begin{equation} \label{MIXING}
\Phi_t = q_1\Phi^{(1)}_t + q_2\Phi^{(2)}_t,
\end{equation}
where $q_{1,2}\geq 0$ and $q_1+q_2=1$. It is easy to show that $\Phi_t$ is, in fact, a CPT map
provided that $\Phi^{(1)}_t$ and $\Phi^{(2)}_t$ are CPT maps. The map $\Phi_t$ will be called
mixture of the maps $\Phi^{(1)}_t$ and $\Phi^{(2)}_t$. This construction can  be extended to an
arbitrary number of dynamical maps $\Phi^{(i)}_t$ in an obvious way
\cite{supp}. To keep the notation simple we will restrict here to the
case of mixtures of two dynamical maps. A simple but well-known example of this construction is 
obtained by taking all maps $\Phi^{(i)}_t$ to be
unitary transformations, in this case the resulting mixture $\Phi_t$ is known as random unitary map 
\cite{Landau1993a,Audenaert2008a}.

\textit{Non-Markovianity.}
To explain the concept of quantum non-Markovianity to be used in the
following \cite{Breuer2009b,Wissman2015a} we consider two
parties, Alice and Bob. Alice prepares a quantum system $S$ in one of two states $\rho_S^1$,
$\rho_S^2$ with probability of $1/2$ each, and then sends the system to Bob. It is Bob's task
to figure out by a single measurement whether the system has been prepared in state $\rho_S^1$
or $\rho_S^2$. It can be shown that by an optimal strategy Bob can find the correct state with
a maximal success probability given by
\begin{equation} \label{P-max}
 P_{\max} = \frac{1}{2} \left(1+ D(\rho_S^1,\rho_S^2)\right),
\end{equation}
where
\begin{equation}\label{trace-distance}
 D(\rho_S^1,\rho_S^2) = \frac{1}{2} ||\rho_S^1-\rho_S^2||
\end{equation}
denotes the trace distance of the quantum states. The trace distance thus represents a
measure for the distinguishability of quantum states. In these relations it is assumed that
Bob receives the quantum system in states $\rho_S^1$ or $\rho_S^2$. However, if we assume
that Alice prepares her states as states of an open system $S$ which is coupled to some
environment $E$, Bob receives instead the states $\rho_S^1(t)=\Phi_t [\rho_S^1]$ or 
$\rho_S^2(t)=\Phi_t [\rho_S^2]$, where $\Phi_t$ denotes the quantum dynamical map
describing the evolution of $S$. The trace distance of the states available to Bob is then
given by $D(\rho_S^1(t),\rho_S^2(t))$ and, hence, the maximal probability with which he can
correctly identify the state is given by
\begin{equation} \label{P-max-Phi}
 P^{\Phi}_{\max}(t) = \frac{1}{2} 
 \left(1+ \frac{1}{2} || \Phi_t [ \rho_S^1 - \rho_S^2 ] || \right).
\end{equation}

A quantum process given in terms of a family of quantum dynamical maps
$\Phi_t$ is defined to be Markovian if the trace distance
$D(\rho_S^1(t),\rho_S^2(t))$ is a monotonically decreasing function of
time $t$ for all pairs of initial states. Hence, quantum
non-Markovianity is characterized by a temporal increase of the trace
distance for a certain pair of initial states. Since the trace
distance represents a measure for the distinguishability of quantum
states, a decrease of the trace distance can be interpreted as a loss
of information from the open system $S$ into the environment $E$.
Correspondingly, any increase of the trace distance corresponds to a
flow of information from the environment back to the open system which
is characteristic of the presence of memory effects. On the basis of
these concepts one can also define a measure for the degree of
non-Markovianity by means of
\begin{equation} \label{NM-measure}
 \mathcal{N}(\Phi) = \max_{\rho^{1,2}_S} 
 \int_{\sigma>0} dt \; \sigma(t),
\end{equation}
where
\begin{equation} \label{trace-distance-deriv}
\sigma(t) \equiv \frac{d}{dt} D\left(\Phi_t\rho^1_S,\Phi_t\rho^2_S\right).
\end{equation}
Thus, $\mathcal{N}(\Phi)$ quantifies the amount of the total information which flows back
from the environment into the open system during the time evolution.

How does this concept of memory effects in quantum mechanics and the associated measure of
non-Markovianity behave under mixing of quantum dynamical maps? It is quite natural to expect
that mixing always makes quantum processes more Markovian. According to several examples 
constructed in the literature \cite{Chruscinski2015a,Kropf2016a} this intuitive expectation is false. 
Indeed, it is even possible that $\Phi_t$ is non-Markovian although the dynamical maps
$\Phi^{(i)}_t$ are Markovian, and even represent quantum dynamical semigroups \cite{supp}. 
Thus, the set of Markovian processes is not convex and the following questions arise: How can memory 
effects emerge through the simple process of mixing quantum processes, and how can this be 
interpreted in terms of a backflow of information from the environment to the open system? 
To discuss these issues we first design an appropriate microscopic description for the mixing procedure.

\begin{figure}[tbh]
\centering
\includegraphics[width=0.45\textwidth]{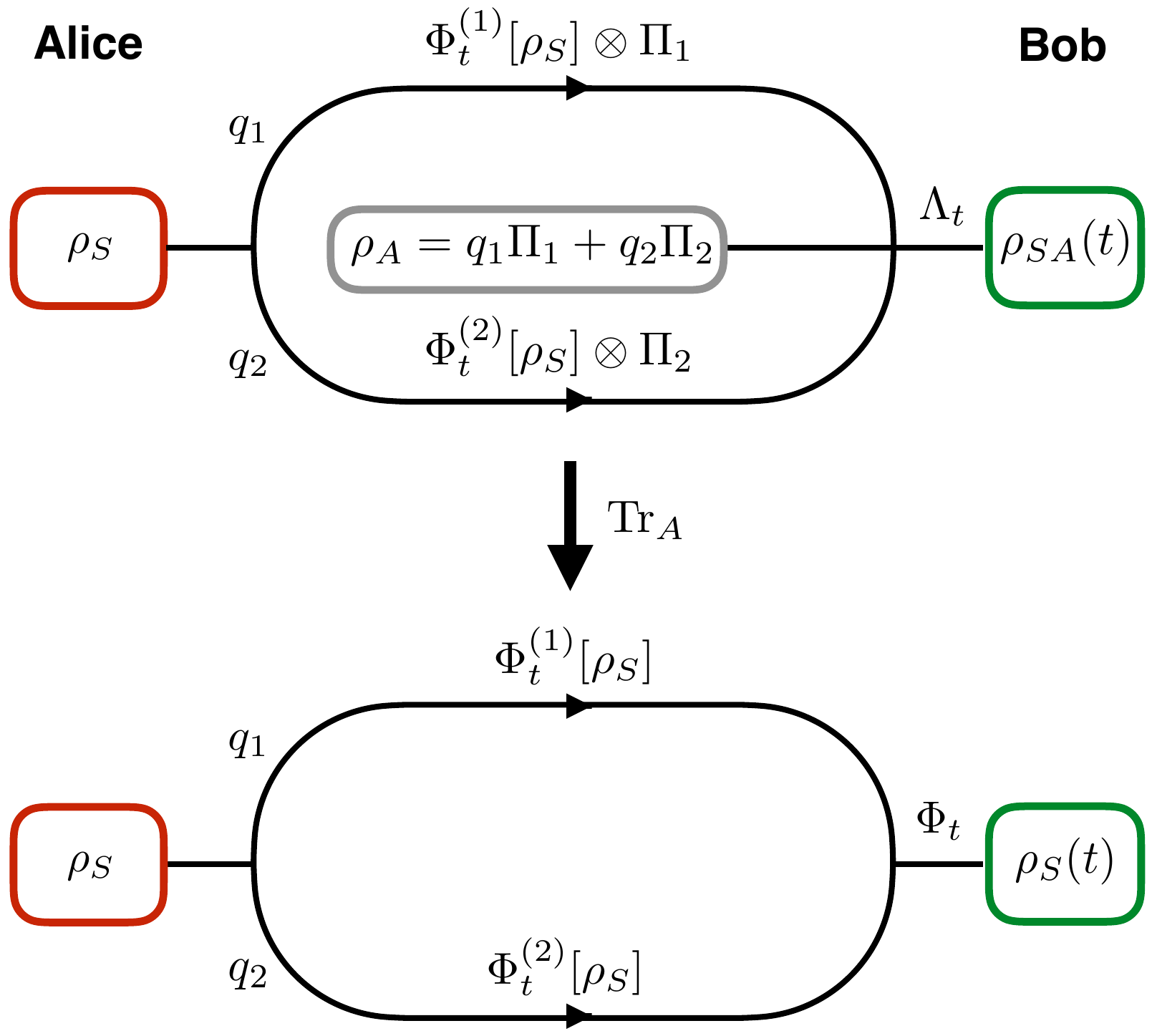}
\caption{Scheme of the microscopic
  interaction leading to the maps $\Lambda_t$ and $\Phi_t$. In both
  cases the system state is coupled to two environments and an ancilla
  in a fixed state. The map $\Lambda_t$ describes the
  state of both system and ancilla at time $t$, while $\Phi_t$
  provides the transformed state of the system
  only.
}%
\label{fig1}
\end{figure}

\textit{Microscopic representation of mixing.}
We start from the following representations for the dynamical maps $\Phi^{(1)}_t$ 
and $\Phi^{(2)}_t$:
\begin{equation}
 \Phi^{(i)}_t [\rho_S] = \Tr_{E_i} \Big[ U^{(i)}_t \rho_S \otimes \rho_{E_i} U^{(i)\dagger}_t \Big],
 \quad i = 1,2.
\label{eq:CPbr}
\end{equation}
These maps act on the density matrices $\rho_S$ of the state space of the open system which we 
denote by $\mathcal{S}(\mathcal{H}_{S})$, where $\mathcal{H}_{S}$ is the underlying Hilbert 
space. For each $i$, $\rho_{E_i}$ is a fixed environmental state taken from the state space 
$\mathcal{S}(\mathcal{H}_{E_i})$ of environment $E_i$, and the unitary time-evolution operator
$U^{(i)}_t$ is taken to be
\begin{equation}
 U^{(i)}_t = \exp (-iH_i t ),
\end{equation}
where $H_i$ is the Hamiltonian for the composite system $S+E_i$ with Hilbert space
$\mathcal{H}_S \otimes \mathcal{H}_{E_i}$. For simplicity we assume the Hamiltonians to be 
time-independent. The generalization to time-dependent Hamiltonians is straightforward.
Finally, $\Tr_{E_i}$ denotes the partial trace over environment $E_i$.

Our goal is to develop a microscopic representation of the time
evolution corresponding to the convex linear combination
\eqref{MIXING}. To this end, we couple our open system $S$ to the two different environments 
$E_1$, $E_2$ and, additionally, to an ancilla system $A$ with a two-dimensional
Hilbert space $\mathcal{H}_A$. The Hilbert space of the total system is thus 
given by 
$\mathcal{H}_S\otimes\mathcal{H}_{E_1}\otimes\mathcal{H}_{E_2}\otimes\mathcal{H}_A$,
and the total system Hamiltonian is taken to be
\begin{equation}
 H =  H_1 \otimes\Pi_1 + H_2\otimes \Pi_2. 
\label{eq:totHammix}
\end{equation}
Here, $H_i$ describes, as above, the coupling between the open system $S$ and environment 
$E_i$, while $\Pi_i=|i\rangle\langle i|$ are orthogonal rank-one projections corresponding to
some basis $|i\rangle$, $i=1,2$, of the ancilla Hilbert space $\mathcal{H}_A$.
Taking as initial state of the ancilla system the fixed state
\begin{equation}
 \rho_A = q_1 \Pi_1 + q_2 \Pi_2,
\label{eq:statoancilla}
\end{equation}
one can introduce a microscopic representation
for the mixture of dynamical maps Eq.~\eqref{MIXING} which is illustrated in Fig.~\ref{fig1}.
Indeed, denoting by 
$U_t=\exp(-iHt)$ the unitary time-evolution operator of the total system,
one can consider the map 
\begin{equation}
 \Lambda_t [ \rho_S ] = \Tr_{E_1} \Tr_{E_2} 
 \big[ U_t \rho_S \otimes \rho_{E_1} \otimes \rho_{E_2} \otimes \rho_A U^{\dagger}_t \big] ,
\label{eq:enlargedmicromixing}
\end{equation}
which can be written as \cite{supp}
\begin{equation}
 \Lambda_t [ \rho_S ] = q_1 \Phi^{(1)}_t [ \rho_S ] \otimes \Pi_1 + q_2 \Phi^{(2)}_t [ \rho_S ] 
 \otimes \Pi_2.
\label{eq:enlargedmicromixingexpr}
\end{equation}
Taking the partial trace with respect to the ancilla degrees of
freedom one obtains from this equation \cite{supp}
\begin{equation}
 \Phi_t [ \rho_S ] = \Tr_{E_1} \Tr_{E_2} \Tr_{A} \big[ U_t \rho_S \otimes \rho_{E_1} \otimes 
 \rho_{E_2} \otimes \rho_A U^{\dagger}_t \big].
\label{eq:micromixing}
\end{equation}
Thus, we have shown that any mixture of quantum dynamical maps of the form of 
Eq.~\eqref{MIXING} admits a microscopic representation of the form \eqref{eq:micromixing}
with the help of an additional ancilla qubit system.
To explain the physical interpretation of this construction we consider again the two parties Alice 
and Bob. Alice prepares the quantum system $S$ in a certain state $\rho_S$ and sends it to Bob 
through quantum channels $\Phi^{(i)}_t$ with respective probabilities $q_i$. Thus, Bob receives the 
states $\Phi^{(i)}_t [ \rho_S ]$ with corresponding probability $q_i$.

Let us suppose first that Bob has access not only to the degrees of freedom of $S$, but also 
to the degrees of freedom of the ancilla system $A$. Bob can then obviously figure out which 
channel has acted on the input state $\rho_S$. This is due to the correlations in the 
system-ancilla state $\Lambda_t [ \rho_S ]$ shown in Eq.~\eqref{eq:enlargedmicromixingexpr}. 
In fact, if Bob measures, for example, $\Pi_1$ he will find $\Pi_1=1$ with probability $q_1$ and in 
this case he knows that the channel $\Phi^{(1)}_t [ \rho_S ]$ has acted on the input state. 
Accordingly, he will get the outcome $\Pi_1=0$ with probability $q_2$ in which case he knows that 
the channel $\Phi^{(2)}_t [ \rho_S ]$ has acted on the input state. Hence, the map $\Lambda_t$ 
describes the situation in which Bob does know the channel which has acted on the input state. 
Note that the ancilla represents, essentially, a classical degree of freedom which does not change 
in time because of $\Tr_{S} \Lambda_t [ \rho_S ] = \rho_A$, and that the correlations between 
system and ancilla are purely classical correlations (no entanglement and zero quantum discord).

\begin{figure}[tbh]
\centering
\includegraphics[width=0.45\textwidth]{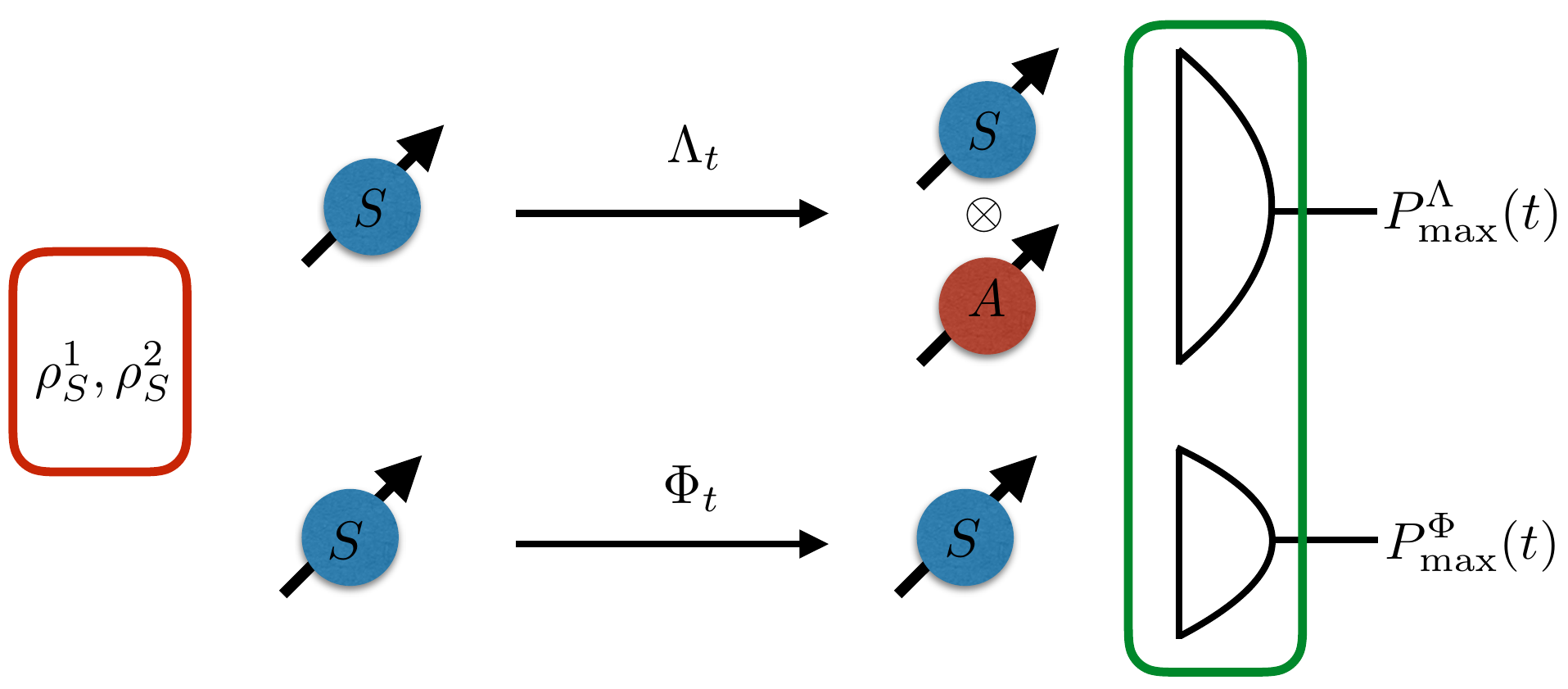}
\caption{Given a pair of initial states
  $\rho_S^{1,2}$ prepared by Alice, Bob has two different optimal
  measurement strategies to discriminate between them for the two
  maps $\Lambda_t$ and $\Phi_t$. In particular $P^{\Lambda}_{\max}(t)
 \geq P^{\Phi}_{\max}(t)$ since in the first case Bob can also access the
  degrees of freedom of the ancilla.
}%
\label{fig2}
\end{figure}

Hence, if Bob has access to the ancilla degree of freedom (see Fig.~\ref{fig2}) the maximal 
success probability with which he can correctly identify the state is given by
\begin{equation}
 p^{\Lambda }_{\textrm{max}} (t) = q_1 p^{ \Phi^{(1)} }_{\textrm{max}} (t) 
 + q_2 p^{ \Phi^{(2)} }_{\textrm{max}} (t).
\end{equation}
Employing expression \eqref{P-max-Phi} and the general relation
\cite{supp}
\begin{equation}
\trn{ \Lambda_t [ X ] } 
=  q_1 \trn{ \Phi^{(1)}_t [ X ] } + q_2 \trn{ \Phi^{(2)}_t [ X ] }
\end{equation}
we obtain
\begin{equation}
p^{\Lambda }_{\textrm{max}} (t) = \frac{1}{2} 
\left( 1 +  \frac{1}{2}\trn{ \Lambda_t [ \rho_S^1-\rho_S^2] } \right).
\label{eq:discriminoto}
\end{equation}
Thus we see that the distinguishability under $\Lambda_t$ is equal to the weighted sum of
the distinguishabilities under $\Phi_t^{(1)}$ and $\Phi_t^{(2)}$. Therefore, if $\Phi_t^{(1)}$
and $\Phi_t^{(2)}$ are Markovian, then $\Lambda_t$ is also Markovian. On the other hand, if
$\Phi_t^{(1)}$ or $\Phi_t^{(2)}$ is non-Markovian, then $\Lambda_t$ can be Markovian or
non-Markovian, depending on whether or not the increase of the trace distance under e.g. 
$\Phi_t^{(1)}$ is compensated by a corresponding decrease of the trace distance under
$\Phi_t^{(2)}$. In this sense on can say that, in general, $\Lambda_t$ is more Markovian
than $\Phi_t^{(1)}$ and $\Phi_t^{(2)}$. Formally, this result can be expressed
by the general relation \cite{supp}
\begin{equation} \label{NM-measure-rel}
 \mathcal{N}(\Lambda) \leq q_1 \mathcal{N}(\Phi^{(1)}) + q_2\mathcal{N}(\Phi^{(2)}).
\end{equation}

\textit{Non-Markovianity induced by mixing.}
Let us now suppose that Bob has no access to the ancilla degrees of
freedom and, hence, 
has no information about whether the channel $\Phi^{(1)}_t$ or $\Phi^{(2)}_t$ has been used by 
Alice. Bob only knows the corresponding probabilities $q_1$ and $q_2$. In this situation he 
has to describe the channel by the convex linear combination \eqref{MIXING} since the states he 
receives are the statistical mixtures $q_1\Phi^{(1)}_t[\rho_S] + q_2\Phi^{(2)}_t[\rho_S]$.
It follows that the maximal probability for correct state identification by Bob is now given by
\begin{equation} \label{P-max-Phi-2}
 P^{\Phi}_{\max}(t) = \frac{1}{2} 
 \left(1+ \frac{1}{2} || \Phi_t [ \rho_S^1 - \rho_S^2 ] || \right).
\end{equation}
Using $\Phi_t = \mathrm{Tr}_A \circ \Lambda_t$ as well as the fact that the trace
operation is a contraction under the trace norm \cite{Ruskai1994a}, we immediately obtain
(see also Fig.~\ref{fig2})
\begin{equation}
 P^{\Phi}_{\max} (t) \leq P^{\Lambda}_{\max} (t)
 = q_1 P^{ \Phi^{(1)} }_{\max} (t) + q_2 P^{ \Phi^{(2)} }_{\max} (t).
\end{equation}
According to this inequality the process $\Phi_t$, in contrast to the process $\Lambda_t$ 
acting on both system and ancilla degree of freedom,
can be non-Markovian even though both $\Phi^{(1)}_t$ and $\Phi^{(2)}_t$ are Markovian. In fact, the inequality 
gives room for the trace distance under $\Phi_t$ to behave non-monotonically although the trace distances under
$\Phi^{(1)}_t$ and $\Phi^{(2)}_t$ are monotonically decreasing corresponding to Markovian
dynamics. The reason for this is obviously the partial trace over the ancilla, which leads to
a loss of information about the channel acting on the system states. Hence, in the case of
non-Markovian dynamics of $\Phi_t$ with $\Phi^{(1)}_t$ and $\Phi^{(2)}_t$ Markovian the
interpretation is that there is a backflow of information from the ancilla $A$ to the open 
system $S$.

\textit{Information backflow analysis.} This idea of information backflow from the ancilla to the system can  be made more precise as follows. We define the quantities
\begin{eqnarray}
 I_{\mathrm{int}}(t) &=& \frac{1}{2} || \Phi_t [ \rho_S^1 - \rho_S^2 ] ||, \\  \label{eq:Iint}
 I_{\mathrm{ext}}(t) &=& \frac{1}{2} || \Lambda_t [ \rho_S^1 - \rho_S^2 ] || -
 \frac{1}{2} || \Phi_t [ \rho_S^1 - \rho_S^2 ] ||. \label{eq:Iext}
\end{eqnarray}
The quantity $I_{\mathrm{int}}(t)$ is the distinguishability in the case that Bob has no access
to the ancilla degrees of freedom. This quantity may thus be interpreted as the internal
information, i.e.~as the information available if only access to the open system $S$ is possible.
On the other hand, the quantity $I_{\mathrm{ext}}(t)$ is the distinguishability including the
ancilla degrees of freedom minus the distinguishability without ancilla. Hence, we can interpret
$I_{\mathrm{ext}}(t)$ as external information, i.e.~as the information which is gained if one
includes the ancilla degrees of freedom. Note that $I_{\mathrm{ext}}(t)\geq 0$ and that
$\Lambda_{t=0}[\rho_S] = \rho_S \otimes \rho_A$ from which it follows that
$I_{\mathrm{ext}}(0)=0$. Moreover, we have
$\mathrm{Tr}_A \Lambda_t[\rho_S]=\Phi_t[\rho_S]$ and
$\mathrm{Tr}_S \Lambda_t[\rho_S]=\rho_A$ which shows that $\Phi_t[\rho_S]\otimes\rho_A$
is the product of the marginals of the state $\Lambda_t[\rho_S]$. 
Now, one can prove the inequality \cite{supp},
\begin{equation} \label{ineq-I_ext}
 I_{\mathrm{ext}}(t) \leq
 D(\Lambda_t[\rho_S^1] , \Phi_t [\rho_S^1] \otimes \rho_A )
 + D(\Lambda_t[\rho_S^2] , \Phi_t [\rho_S^2] \otimes \rho_A ),
\end{equation}
where the quantity $D(\Lambda_t[\rho_S] ,\Phi_t [\rho_S] \otimes \rho_A)$,
representing the trace distance between the state $\Lambda_t[\rho_S]$ and the product
of its marginals, provides a measure for the system-ancilla correlations in the state
$\Lambda_t[\rho_S]$. We recall that these correlations are of purely classical nature.

The inequality \eqref{ineq-I_ext} shows that the external information is bounded from above by the 
sum of the correlation measures of the states $\Lambda_t[\rho_S^1]$ and $\Lambda_t[\rho_S^2]$. 
In particular, when $I_{\mathrm{ext}}(t)$ starts to increase over the initial value 
$I_{\mathrm{ext}}(0)=0$ correlations between the open system and the ancilla are created.
In other words, any nonzero external information implies that there are system-ancilla
correlations which are inaccessible to Bob if he can only measure the observables of the
open system $S$. 

For the interesting special case referred to above, namely that
$\Phi^{(1)}_t$ and $\Phi^{(2)}_t$ are Markovian while the convex mixture
$\Phi_t$ is non-Markovian, the trace distance
$\frac{1}{2} || \Lambda_t [ \rho_S^1 - \rho_S^2 ] ||$ decreases
monotonically, so that we have
\begin{equation} \label{ineq-infoflow}
 \dot{I}_{\mathrm{int}}(t) + \dot{I}_{\mathrm{ext}}(t) \leq 0.
\end{equation}
However, the quantity
$I_{\mathrm{int}}(t) = \frac{1}{2} || \Phi_t [ \rho_S^1 - \rho_S^2 ] ||$
must be a non-monotonic function of time $t$ for a certain pair of
initial states $\rho_S^{1,2}$.  Hence, it follows that
$\dot{I}_{\mathrm{int}}(t)>0$ for certain times $t$ which implies
$\dot{I}_{\mathrm{ext}}(t)<0$, i.e. there is a nonzero backflow of information
from the ancilla into the open system. This clearly supports our
interpretation of mixing-induced non-Markovianity. We note
that for $\dot{I}_{\mathrm{int}}(t)<0$ we cannot generally draw any
conclusion about the sign of the external information
$\dot{I}_{\mathrm{ext}}(t)$ from inequality
\eqref{ineq-infoflow}. This is due to the fact that the open system
can lose information both to the ancilla {\textit{and}} to the environments
$E_i$. Finally, we consider the particularly relevant case of a random 
mixture of unitary maps. In this case the trace distance
$\frac{1}{2} || \Lambda_t [ \rho_S^1 - \rho_S^2 ] ||$ is constant in time and, hence,
the inequality of Eq.~\eqref{ineq-infoflow} actually becomes an
equality, corresponding to the fact that now the system can lose information only to the ancilla.

\textit{Conclusions.}  We have constructed a microscopic representation for a quantum 
dynamical map arising as a convex mixture of dynamical maps. Our construction allows to 
understand the relationship between the Markovianity of the quantum dynamical map
obtained by mixing and the Markovianity of the single elements of the mixture. The 
analysis shows in particular that counterintuitive behaviours, such as the emergence of 
non-Markovianity by mixing Markovian semigroups or unitary dynamics, can be clearly 
explained and understood within a consistent characterization of non-Markovianity in terms of the 
flow of information between the open system and its environment. 

The crucial point of our construction is the fact that the operation of mixing 
involves an ancilla system which behaves essentially as a classical degree of freedom,
acting as a \textit{random number generator} which determines the choice of the quantum channel. 
It therefore plays a similar role as the classical device considered in the seminal work on quantum 
correlations \cite{Werner1989}.
While the reduced state of the ancilla does not change in time, correlation between the
open system and the ancilla are built up during the time evolution. Thus, the open system can
exchange information with the ancilla degree of freedom by virtue of these correlations, and
it is this exchange of information which leads to mixing-induced quantum non-Markovianity.
These results clearly reinforce the physical motivation underlying the description of quantum 
memory in terms of distinguishability of quantum states, system-environment correlations 
and information flow between system and environment.

\acknowledgments HPB and BV acknowledge support from the European Union (EU)
through the Collaborative Project QuProCS (Grant Agreement No. 641277).

\appendix

\section{Proofs}

\subsection{Proof of Eqs.~(12) and (13)}

In order to prove Eq.~(12) of the main text we start from the definition
\begin{equation}
 \Lambda_t [ \rho_S ] = \Tr_{E_1} \Tr_{E_2} 
 \big[ U_t \rho_S \otimes \rho_{E_1} \otimes \rho_{E_2} \otimes \rho_A U^{\dagger}_t \big].
\end{equation}
Note that this is a CPT map
\begin{equation}
 \Lambda_t: \mathcal{S}(\mathcal{H}_S) \rightarrow 
 \mathcal{S}(\mathcal{H}_{S} \otimes \mathcal{H}_A).
\end{equation}
Since $\Pi_1 \Pi_2 = 0$ we can split the unitary time-evolution operator as
\begin{equation}
 U_t = e^{ - i H t } =  e^{ - i H_1 \Pi_1 t }  e^{ - i H_2  \Pi_2 t }.
\label{eq:globmixunit}
\end{equation}
Using also Eq.~(10) of the main text we find
\begin{eqnarray}
\Lambda_t [ \rho_S ] &=& \textrm{ } q_1 \Tr_{E_1} \Tr_{E_2} \big[ U_t \rho_S \otimes \rho_{E_1} 
\otimes \rho_{E_2} \otimes \Pi_1 U^{\dagger}_t \big] \nonumber \\
&& + \textrm{ }  q_2 \Tr_{E_1} \Tr_{E_2} \big[ U_t \rho_S \otimes \rho_{E_1} \otimes \rho_{E_2} 
\otimes \Pi_2 U^{\dagger}_t \big] \nonumber \\
&=& \textrm{ }  q_1 \Tr_{E_1} \Tr_{E_2} \big[ U^{(1)}_t \rho_S \otimes \rho_{E_1} \otimes 
\rho_{E_2} 
\otimes \Pi_1 U^{(1)\dagger}_t \big] \nonumber \\
&& + \textrm{ }  q_2 \Tr_{E_1} \Tr_{E_2} \big[ U^{(2)}_t \rho_S \otimes \rho_{E_1} \otimes 
\rho_{E_2} 
\otimes \Pi_2 U^{(2)\dagger}_t \big] \nonumber \\
&=& \textrm{ }  q_1 \Tr_{E_1} \big[ U^{(1)}_t \rho_S \otimes \rho_{E_1}  U^{(1)\dagger}_t \big] 
\otimes \Pi_1 \nonumber \\
&& + \textrm{ }  q_2 \Tr_{E_2} \big[ U^{(2)}_t \rho_S \otimes \rho_{E_2} U^{(2)\dagger}_t \big]  
\otimes \Pi_2.
\end{eqnarray}
Employing Eq.~(7) of the main text we can rewrite this as
\begin{equation}
 \Lambda_t [ \rho_S ] = q_1 \Phi^{(1)}_t [ \rho_S ] \otimes \Pi_1 + q_2 \Phi^{(2)}_t [ \rho_S ] 
 \otimes \Pi_2
\label{eq:enlargedmicromixingexpr-2}
\end{equation}
which proves Eq.~(12) of the main text. Tracing over the ancilla degree of freedom we find
\begin{eqnarray}
 \Tr_A \Lambda_t [ \rho_S ] &=& q_1 \Phi^{(1)}_t [ \rho_S ]  
 + q_2 \Phi^{(2)}_t [ \rho_S ] \nonumber \\
 &=& \Phi_t [ \rho_S ] \\
 &=&  \Tr_{E_1} \Tr_{E_2} \Tr_{A} 
 \big[ U_t \rho_S \otimes \rho_{E_1} \otimes \rho_{E_2} \otimes \rho_A U^{\dagger}_t \big],
 \nonumber
\label{eq:rellambdaphi}
\end{eqnarray}
which proves Eq.~(13) of the main text.

\subsection{Proof of Eq.~(15)}

To prove Eq.~(15) of the main text we start from
\begin{equation}
\Lambda_t [ X ] = q_1 \Phi^{(1)}_t  [ X ] \otimes \Pi_1 + q_2 \Phi^{(2)}_t  [ X ] \otimes \Pi_2,
\end{equation} 
where $X$ is any system operator. Note that the operators $ \Phi^{(1)}_t  [ X ] \otimes \Pi_1 $ 
and $ \Phi^{(2)}_t  [ X ] \otimes \Pi_2 $ have orthogonal support in 
$ \mathcal{H}_{S} \otimes \mathcal{H}_A $ and, hence, we have
\begin{equation} 
 || \Lambda_t [ X] || = q_1 || \Phi^{(1)}_t  [ X ] \otimes \Pi_1 || 
 +  q_2 || \Phi^{(2)}_t  [ X ] \otimes \Pi_2 || .
\end{equation}
Using $ || A \otimes B || = || A || \cdot|| B || $ and $||\Pi_{1,2}||=1$ we obtain Eq.~(15)
of the main text.

\subsection{Proof of Eq.~(17)}

To prove Eq.~(17) of the main text we use the definition of the non-Markovianity measure
for the map $\Lambda_t$:
\begin{equation} \label{eq:def-N}
 {\mathcal{N}}(\Lambda) = \max_{\rho_S^{1,2}} 
 \int_{\sigma_{\Lambda}>0} dt \; \sigma_{\Lambda} (t) ,
\end{equation}
where
\begin{equation}
 \sigma_{\Lambda} (t) \equiv \frac{1}{2} \frac{d}{dt} || \Lambda_t [ \rho_S^1 - \rho_S^2 ] || .
\end{equation}
Let $\rho_S^{1,2}$ be an optimal state pair for which the maximum in Eq.~\eqref{eq:def-N} is
attained. Then we can write
\begin{equation} \label{eq:def-N-2}
 {\mathcal{N}}(\Lambda) = \int_{\sigma_{\Lambda}>0} dt \; \sigma_{\Lambda} (t).
\end{equation}
Using Eq.~(15) of the main text we get
\begin{equation}
 \sigma_{\Lambda} (t) = q_1 \sigma_{1} (t) + q_2 \sigma_{2} (t),
\label{eq:rel_sigma_gen}
\end{equation}
where
\begin{equation}
 \sigma_{i} (t) \equiv \frac{1}{2} \frac{d}{dt}  || \Phi^{(i)}_t [ \rho_S^1 - \rho_S^2 ] || .
\label{eq:ref_sigma_i}
\end{equation}
Thus, we find
\begin{equation} \label{eq:def-N-3}
 {\mathcal{N}}(\Lambda) = q_1 \int_{\sigma_{\Lambda}>0} dt \; \sigma_1 (t) 
 + q_2 \int_{\sigma_{\Lambda}>0} dt \; \sigma_2 (t). 
\end{equation}
Now, we have
\begin{equation} \label{ineq-N1}
 \int_{\sigma_{\Lambda}>0} dt \; \sigma_{1} (t) 
 \leq  \int_{\sigma_{1}>0} dt \; \sigma_{1} (t).
\end{equation}
This is due to the fact that the integration on the right-hand side is extended over
all regions in which the function $\sigma_1(t)$ is positive and, hence, the integral
on the right-hand side is larger or equal to the integral of $\sigma_1(t)$ over any
other region, in particular over the region given by $\sigma_{\Lambda}(t)>0$.
Moreover, we obtain directly from the definition of quantum non-Markovianity that
\begin{equation}
 \int_{\sigma_{1}>0} dt \; \sigma_{1} (t) \leq  {\mathcal{N}} ( \Phi^{(1)} ).
 \label{eq:more_N_i}
\end{equation}
Together with \eqref{ineq-N1} this yields
\begin{equation} \label{ineq-1}
 \int_{\sigma_{\Lambda}>0} dt \; \sigma_{1} (t) \leq  {\mathcal{N}} ( \Phi^{(1)} ).
\end{equation}
In the same manner we obtain
\begin{equation} \label{ineq-2}
 \int_{\sigma_{\Lambda}>0} dt \; \sigma_{2} (t) \leq  {\mathcal{N}} ( \Phi^{(2)} ).
\end{equation}
Using Eqs.~\eqref{ineq-1} and \eqref{ineq-2} in Eq.~\eqref{eq:def-N-3} we finally get the
desired result:
\begin{equation} \label{NM-measure-rel-2}
 \mathcal{N}(\Lambda) \leq q_1 \mathcal{N}(\Phi^{(1)}) + q_2\mathcal{N}(\Phi^{(2)}).
\end{equation}

\subsection{Proof of Eq.~(22)}

To prove Eq.~(22) of the main text we start from the definition
\begin{equation} 
 I_{\textrm{ext}} (t) = \frac{1}{2}|| \Lambda_t [ \rho_S^1 - \rho_S^2 ]  || 
 - \frac{1}{2} || \Phi_t [ \rho_S^1 - \rho_S^2 ]  ||. 
\label{eq:Iext_gen}
\end{equation}
Since $ \rho_A \in \mathcal{S} ( \mathcal{H}_A ) $ has unit trace norm we have
\begin{equation}
|| \Phi_t [ \rho_S^1 - \rho_S^2 ]  || = || \Phi_t [ \rho_S^1 - \rho_S^2 ] 
\otimes \rho_A ||.
\end{equation}
Using the triangular inequality for the trace norm,
\begin{equation}
 \Big| || A || - || B ||  \Big|  \leq || A - B ||,
\end{equation}
we obtain
\begin{eqnarray}
 I_{\textrm{ext}} (t) &\leq &  \frac{1}{2} 
 ||  \Lambda_t [ \rho_S^1 - \rho_S^2 ]  -  \Phi_t [ \rho_S^1 - \rho_S^2 ]  
 \otimes \rho_A  || \nonumber \\
 &=& \frac{1}{2} || ( \Lambda_t [ \rho_S^1 ] - \Phi_t [ \rho_S^1 ] \otimes \rho_A ) \nonumber \\
 && \quad - ( \Lambda_t[ \rho_S^2 ]  - \Phi_t [ \rho_S^2 ] \otimes \rho_A  ) ||.
\end{eqnarray} 
Employing the triangular inequality 
\begin{equation}
 || A - B || \leq || A || + || B ||
\end{equation}
we finally get Eq.~(22) of the main text:
\begin{eqnarray}
 I_{\textrm{ext}} (t) &\leq& \frac{1}{2} 
 || \Lambda_t [ \rho_S^1 ] - \Phi_t [ \rho_S^1 ] \otimes \rho_A || \nonumber \\
&& + \frac{1}{2} || \Lambda_t[ \rho_S^2 ]  - \Phi_t [ \rho_S^2 ] \otimes \rho_A ||.
\end{eqnarray}

\section{MIXING QUANTUM PROCESSES}

For simplicity, the presentation of the main text has been restricted to the mixing of two
dynamical maps. Our results can easily be generalized to an arbitrary number $n$ of
dynamical maps $\Phi_t^{(i)}$, where $i=1,2,\ldots,n$. The convex mixture of such maps
is defined by
\begin{equation}
\Phi_t = \sum_{i=1}^n q_i  \Phi^{(i)}_t,
\label{eq:rellambdaphi_gen}
\end{equation}
where $q_i$ is a probability distribution, i.e.~$q_i\geq 0$ and $\sum_i q_i = 1$. In order to
construct the corresponding microscopic representation we take an ancilla $A$ with
$n$-dimensional Hilbert space $\mathcal{H}_A$. Introducing an orthonormal basis
$\vert i \rangle$ in this space we define rank-one projection operators
$ \Pi_i = \vert i \rangle \langle i \vert $ and a fixed initial state of the ancilla system
\begin{equation}
 \rho_A = \sum_{i=1}^n q_i \Pi_i . 
\label{eq:statoancilla_gen}
\end{equation}
The Hamiltonian of the total system is taken to be
\begin{equation}
  H = \sum_{i=1}^n H_i \otimes \Pi_i,
\label{eq:totHammix_gen}
\end{equation}
where $H_i$ describes the interaction of the open system $S$ with environment $E_i$.
The time evolution operator factorizes,
\begin{equation}
 U_t = \exp (- i H t) = \prod_{i=1}^n \exp (- i H_i \Pi_i t) ,
\label{eq:globmixunit_gen}
\end{equation}
because of $\Pi_i\Pi_j = \delta_{ij} \Pi_i$. The map defined by
\begin{equation}
 \Lambda_t [ \rho_S] = \Tr_{E_1} ... \Tr_{E_n } \Big[ U_t \rho_S \otimes \rho_{E_1} \otimes 
 \ldots \otimes \rho_{E_n} \otimes \rho_A U^{\dagger}_t \Big] 
\label{eq:enlargedmicromixing_gen}
\end{equation}
can thus be written as
\begin{equation}
\Lambda_t [ \rho_S ] = \sum_{i=1}^n q_i \Phi^{(i)}_t [ \rho_S] \otimes \Pi_i.
\label{eq:enlargedmicromixingexpr_gen}
\end{equation}
Taking the partial trace over the ancilla degrees of freedom we find
\begin{eqnarray}
 \Tr_A \Lambda_t [ \rho_S ] &=& \sum_{i=1}^n q_i \Phi^{(i)}_t [ \rho_S] \nonumber \\
 &=& \Phi_t [ \rho_S ] \\
 &=& \Tr_{E_1} ... \Tr_{E_n } \Tr_A \Big[ U_t \rho_S \otimes \rho_{E_1}
 \ldots \rho_{E_n} \otimes \rho_A U^{\dagger}_t \Big] 
 \nonumber 
\end{eqnarray}
which is the desired microscopic representation of the mixture $\Phi_t$ 
analogous to Eq.~(13) of the main text.

\section{GENERALIZED NON-MARKOVIANITY}

In the main text we have used the concept of quantum non-Markovianity based on the trace
distance which represents a measure for the distinguishability of quantum states 
$\rho_S^1$ and $\rho_S^2$ prepared with equal probabilities of $1/2$. Recently, this concept of
non-Markovianity has been extended to include the case that $\rho_S^1$ and $\rho_S^2$ 
are prepared with different probabilities $p_1$ and $p_2$ \cite{Wissman2015a,Breuer2016a}.
It can be shown that by an optimal strategy Bob can now distinguish the states with a maximal
probability given by
\begin{equation}
 P_{\textrm{max}}^{\Phi} (t) = \frac{1}{2} \left(1+  || \Phi_t [ \Delta ] || \right) , 
\end{equation}
where 
\begin{equation}
\Delta = p_1 \rho_S^1 - p_2 \rho_S^2
\end{equation}
is the Helstrom matrix \cite{Helstrom1976}. 
Consequently, the generalized measure of non-Markovianity is defined by means of
\begin{equation}
\tilde{\mathcal{N}}(\Phi) = \max_{p_i,\rho^{i}_S}  \int_{\tilde{\sigma}>0} dt \; \tilde{\sigma}(t),
\label{NM-measure_gen}
\end{equation}
where
\begin{equation}
\tilde{\sigma}(t) \equiv \frac{d}{dt}  || \Phi_t [ \Delta ] ||.
\label{trace-distance-deriv_gen}
\end{equation}
The discussion presented in the main text can be generalized straightforwardly to this
more general notion of quantum non-Markovianity. For example, it can be shown that
Eq.~(17) of the main text becomes
\begin{equation} \label{NM-measure-rel-new}
 \tilde{\mathcal{N}}(\Lambda) \leq q_1 \tilde{\mathcal{N}}(\Phi^{(1)}) 
 + q_2 \tilde{\mathcal{N}}(\Phi^{(2)}),
\end{equation}
while Eq.~(22) of the main text now takes the form
\begin{eqnarray} 
 I_{\mathrm{ext}}(t) &\leq&
 p_1 || \Lambda_t[\rho_S^1] - \Phi_t [\rho_S^1] \otimes \rho_A || \nonumber \\
 && + p_2 || \Lambda_t[\rho_S^2] - \Phi_t [\rho_S^2] \otimes \rho_A ||,
\end{eqnarray}
where $I_{\textrm{ext}}(t) \equiv || \Lambda_t [ \Delta ]  || - || \Phi_t [ \Delta ] ||$.

\section{EXAMPLES}

\subsection{Non-Markovian mixtures of semigroups}

\begin{figure}[tbh]
\centering
\includegraphics[width=80mm]{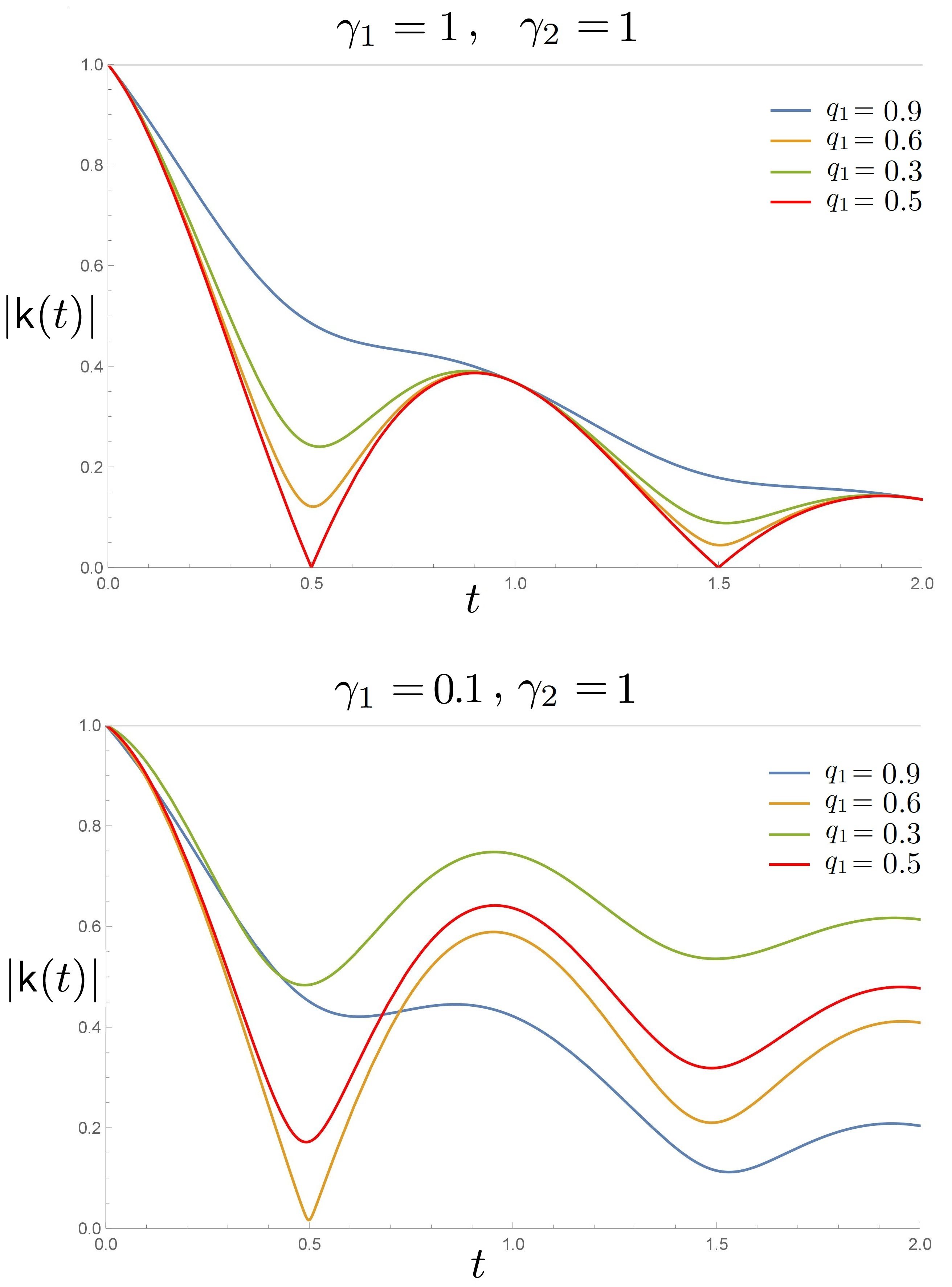}
\caption{The distinguishability (\ref{eq:osckappa}) of optimal state pairs as a function of time.
In these graphs $ \lambda_1 = 2 \pi $ and $\lambda_2 = 0 $.}
\label{fig:subgigante1_mixing}
\end{figure}

\begin{figure}[tbh]
\centering
\includegraphics[width=80mm]{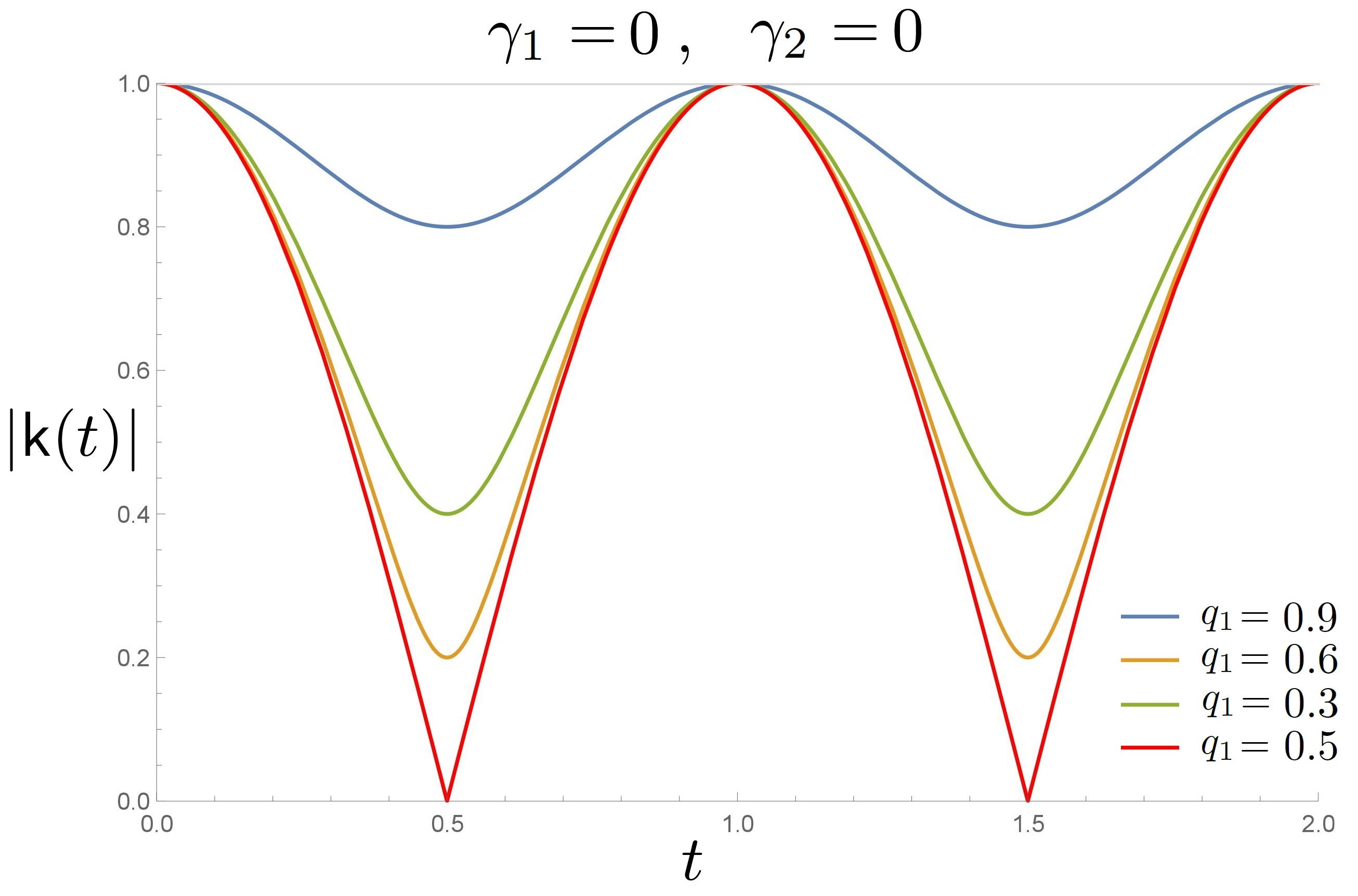}
\caption{The distinguishability (\ref{eq:osckappa}) of optimal state pairs as a function of time
for $ \lambda_1 = 2 \pi$ and $\lambda_2 = 0 $, in the case $ \gamma_1 = \gamma_2 = 0 $
corresponding to a random unitary map.}
\label{fig:subgigante2_mixing_soloUnit}
\end{figure}

For a qubit system consider for $k=1,2$ the maps
\begin{equation}
\Phi^{(k)}_t [ \rho_S ] = \mu_k (t) \rho_S + ( 1 - \mu_k (t) )
\sum_{j=0,1} Q_j  \rho_S Q_j,
\label{eq:procexWis}
\end{equation}
where $ Q_j =  | j \rangle \langle  j |$ are the
projections onto the eigenstates of the operator $\sigma_z$ for the
system and the complex coefficients $\mu_k (t)$ are given by
\begin{equation}
\mu_k (t) = e^{ - ( \gamma_k + i \lambda_k) t}.
\label{eq:muexwis}
\end{equation}
These maps describe pure dephasing of the qubit, obeying a
Markovian semigroup dynamics with Hamiltonian contribution
$({\lambda_k }/{2}) \sigma_z$, and a single Lindblad operator $\sigma_z$ with
corresponding rate $\gamma_k$. Considering a convex mixture of $\Phi^{(1)}_t$
and $\Phi^{(2)}_t$ as in Eq.~(1) of the main text one has that coherences 
evolve as 
 \begin{equation}
  \label{eq:1}
   \langle 1 \vert \rho_S(t)  \vert 0 \rangle = \mathsf{k} (t)\langle 1 \vert \rho_S  \vert 0 \rangle,
 \end{equation}
where
\begin{equation}
  \label{eq:2}
  \mathsf{k} (t)=q_1 \mu_1 (t) + q_2 \mu_2 (t).
\end{equation}
The distinguishability of an optimal pair of states (a pair of states for which
the maximum in Eq.~(5) of the main text is attained) is given by the modulus of
$\mathsf{k} (t)$,
\begin{eqnarray}
 \lefteqn{ | \mathsf{k} (t) | }  && \label{eq:osckappa} \\
 &=&  \sqrt{ q_1^2 e^{-2 \gamma_1 t } + q_2^2 e^{-2 \gamma_2 t } 
 + 2 q_1 q_2 e^{- ( \gamma_1 + \gamma_2 ) t } \cos ( \Delta_{ \lambda} t ) }. \nonumber
\end{eqnarray} 
For the case $ \Delta_\lambda\!=\! \lambda_2 - \lambda_1 \not =0$ this distinguishability can
indeed exhibit a non-monotonic behavior, corresponding to a backflow
of information, even though both $\Phi^{(1)}_t$ and $\Phi^{(2)}_t$ describe a Markovian dynamics.
Examples are shown in Fig.~\ref{fig:subgigante1_mixing} and
Fig.~\ref{fig:subgigante2_mixing_soloUnit}. Note in particular that for the special case 
$\gamma_1=\gamma_2=0$ one recovers an example of random unitary map.

\subsection{Information flow analysis}

\begin{figure}[tbh]
\centering
\includegraphics[width=75mm]{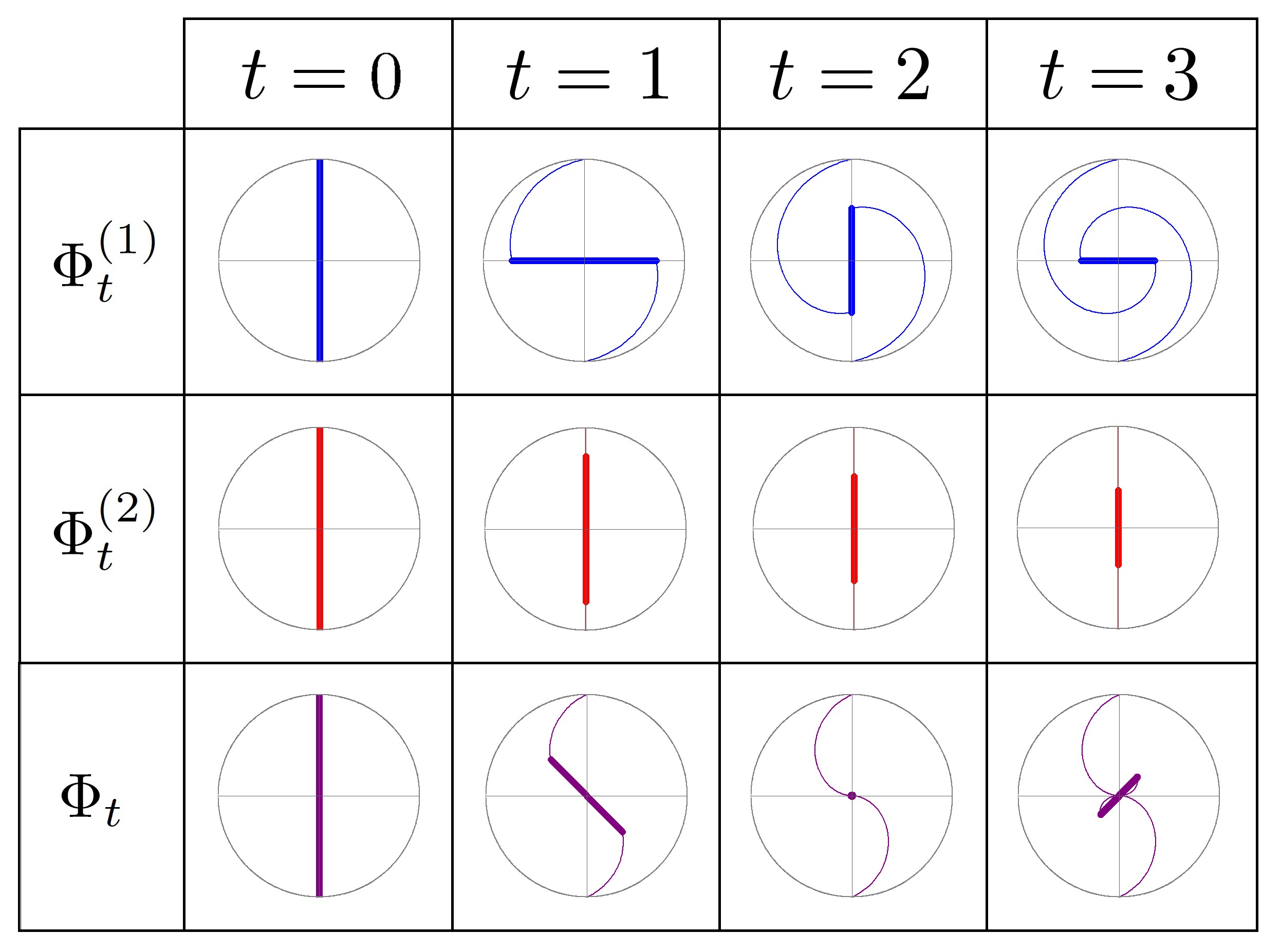}
\caption{Dynamics of the initial states (\ref{eq:myStates}) under the maps 
$ \Phi_t^{(1)} $, $ \Phi_t^{(2)} $ and $ \Phi_t $. 
The thin lines show the trajectories of the states, while the bold straight lines
represent the trace distance at the given time. 
The semigroups $ \Phi_t^{(1)} $ and $ \Phi_t^{(2)} $ yield a monotonically decreasing 
distinguishability, while we obtain revivals of the distinguishability for the convex mixture 
$\Phi_t = \frac{1}{2} [\Phi_t^{(1)} + \Phi_t^{(2)} ] $. In fact, the value of 
distinguishability decreases and reaches zero for 
$ t= 2 $, as $ \Phi_{t=2} [ \rho_S^1 ] = \Phi_{t=2} [ \rho_S^2] $, and then grows back to positive 
values at later times.}
\label{fig:new_tabella}
\end{figure}

\begin{figure}[tbh]
\centering
\includegraphics[width=75mm]{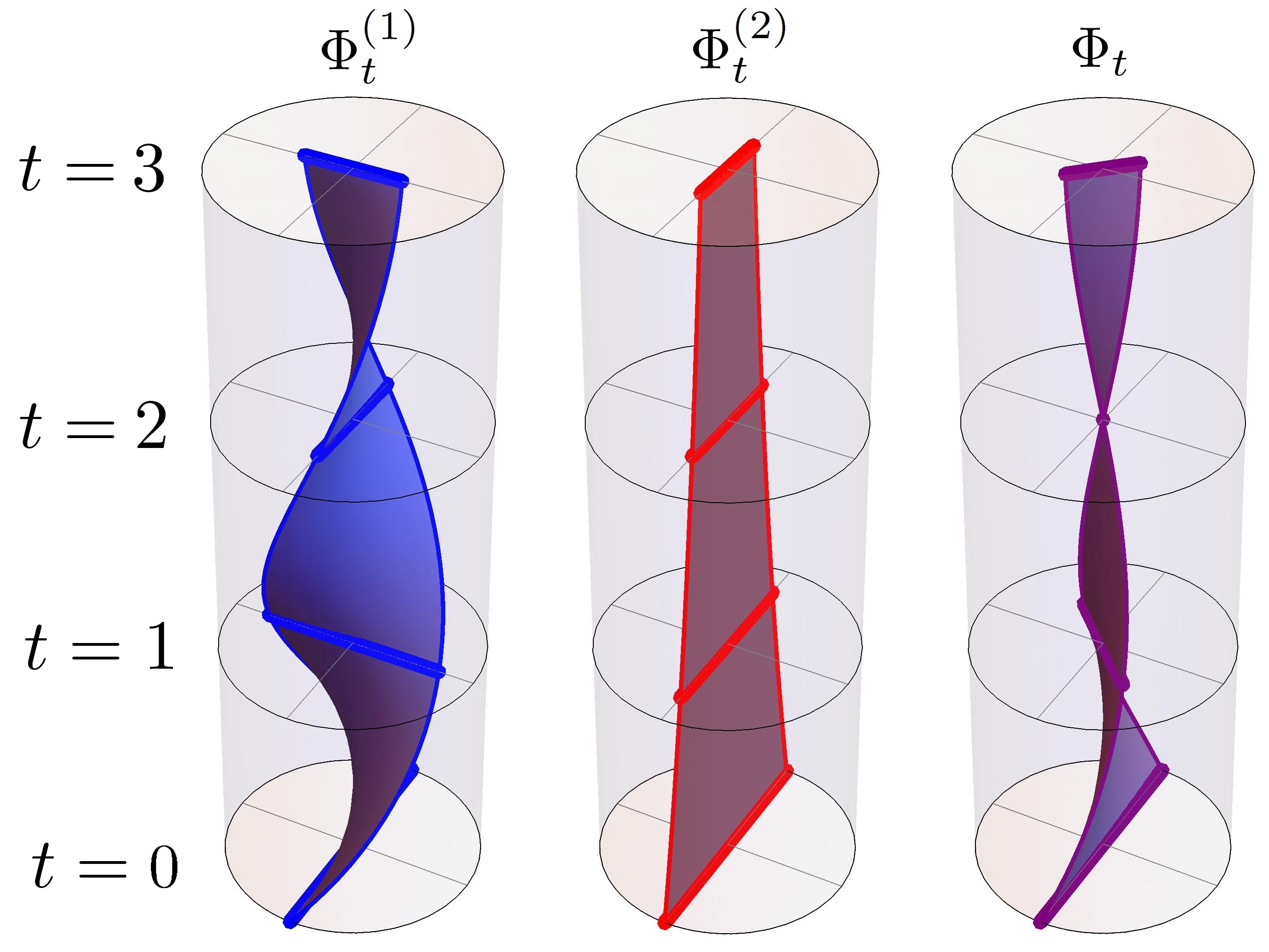}
\caption{Visualization of the dynamics of the initial pair of states (\ref{eq:myStates}), together with 
their trace distance under the maps $ \Phi_t^{(1)} $, $ \Phi_t^{(2)} $ and $ \Phi_t $. 
As in Fig.~\ref{fig:new_tabella} we indicate the trace distance by straight bold lines at times
$t=0,1,2,3$.}
\label{fig:elicheFinal}
\end{figure}

To further illustrate the dynamics let us consider equal weights $ q_1 = q_2 = 1/2 $ in the convex 
mixture of dynamical processes, so that $\Phi_t$ is simply given by the average 
of $ \Phi_t^{(1)}$ and $\Phi_t^{(2)}$. We choose 
$ \gamma_1 = \gamma_2 = 1/3 $ and $ \lambda_1 = \pi/2 $, $ \lambda_2 = 0 $ and, as initial 
states, the orthogonal pair of states
\begin{equation}
 \rho_S^1 = \frac{I_S + \sigma_S^2 }{2} , \qquad
 \rho_S^2 = \frac{I_S - \sigma_S^2 }{2} , \label{eq:myStates}
\end{equation}
which evolve in the equatorial plane of the Bloch sphere. In Figs.~\ref{fig:new_tabella} and 
\ref{fig:elicheFinal} we visualize the dynamics under the various maps and how the mixing process 
leads to non-Markovian dynamics.

\begin{figure}[tbh]
\centering
\includegraphics[width=0.4\textwidth]{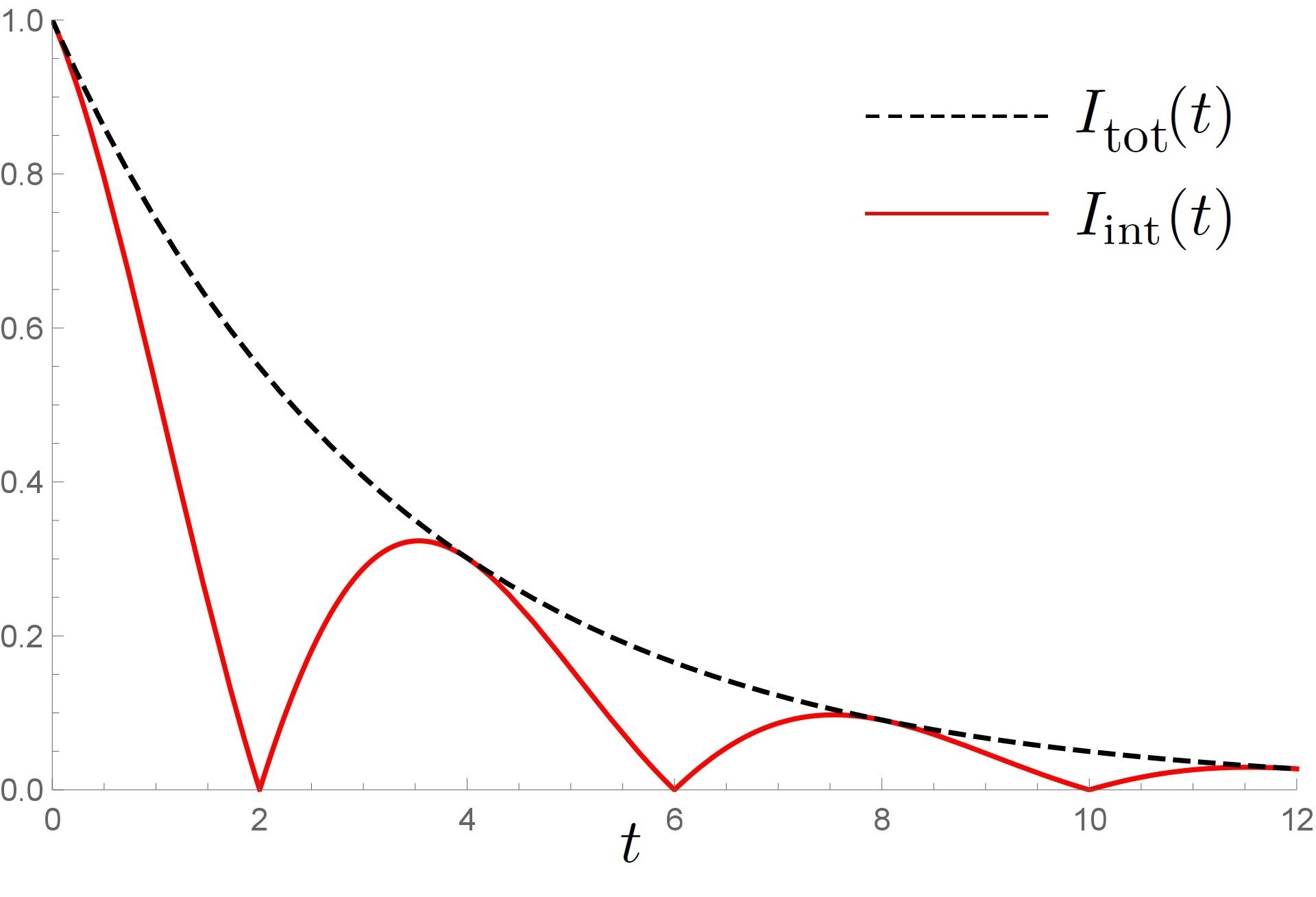}
\caption{Internal and total information as functions of time.
The internal information (\ref{eq:intInfoEsempio}) oscillates and is bounded by the total information 
(\ref{eq:totInfoEsempio}). 
Note the decrease of the total information, signalling the loss of information towards the dissipative 
environments generating the single semigroups $ \Phi_t^{(1)} $ and $ \Phi_t^{(2)} $.}
\label{fig:ITdashed}
\end{figure}

Let us analyze the flow of information by means of the quantities (see Eqs.~(20) and (21) 
of the main text)
\begin{eqnarray}
 I_{\textrm{int}} (t) &=& \frac{1}{2} || \Phi_t [ \rho_S^1  - \rho_S^2 ] ||, \label{eq:intInfo} \\
 I_{\textrm{tot}} (t) &=& I_{\textrm{int}} (t) + I_{\textrm{ext}} (t)
 = \frac{1}{2} || \Lambda_t [ \rho_S^1 - \rho_S^2 ] ||. \label{eq:totInfo}
\end{eqnarray}
For the choice (\ref{eq:myStates}) the internal information is given by 
$\vert \mathsf{k} (t) \vert$ which reads in this specific case
\begin{equation}
 I_{\textrm{int}} (t) = e^{-t/3} |\cos(\pi t/4)|, 
 \label{eq:intInfoEsempio}
\end{equation}
while using Eq. (15) of the main text we obtain
\begin{eqnarray}
 I_{\textrm{tot}} (t) &=& 
 \frac{1}{2} || \Phi^{(1)}_t [ \rho_S^1  - \rho_S^2 ] || + \frac{1}{2} || \Phi^{(2)}_t [ \rho_S^1  - 
 \rho_S^2 ] || \nonumber \\
 &=& \frac{1}{2} e^{ - \gamma_1 t } + \frac{1}{2} e^{ - \gamma_2 t } = e^{-t/3} .
\label{eq:totInfoEsempio}
\end{eqnarray}
These expressions are plotted in Fig.~\ref{fig:ITdashed}.

\begin{figure}[tbh]
\centering
\includegraphics[width=0.4\textwidth]{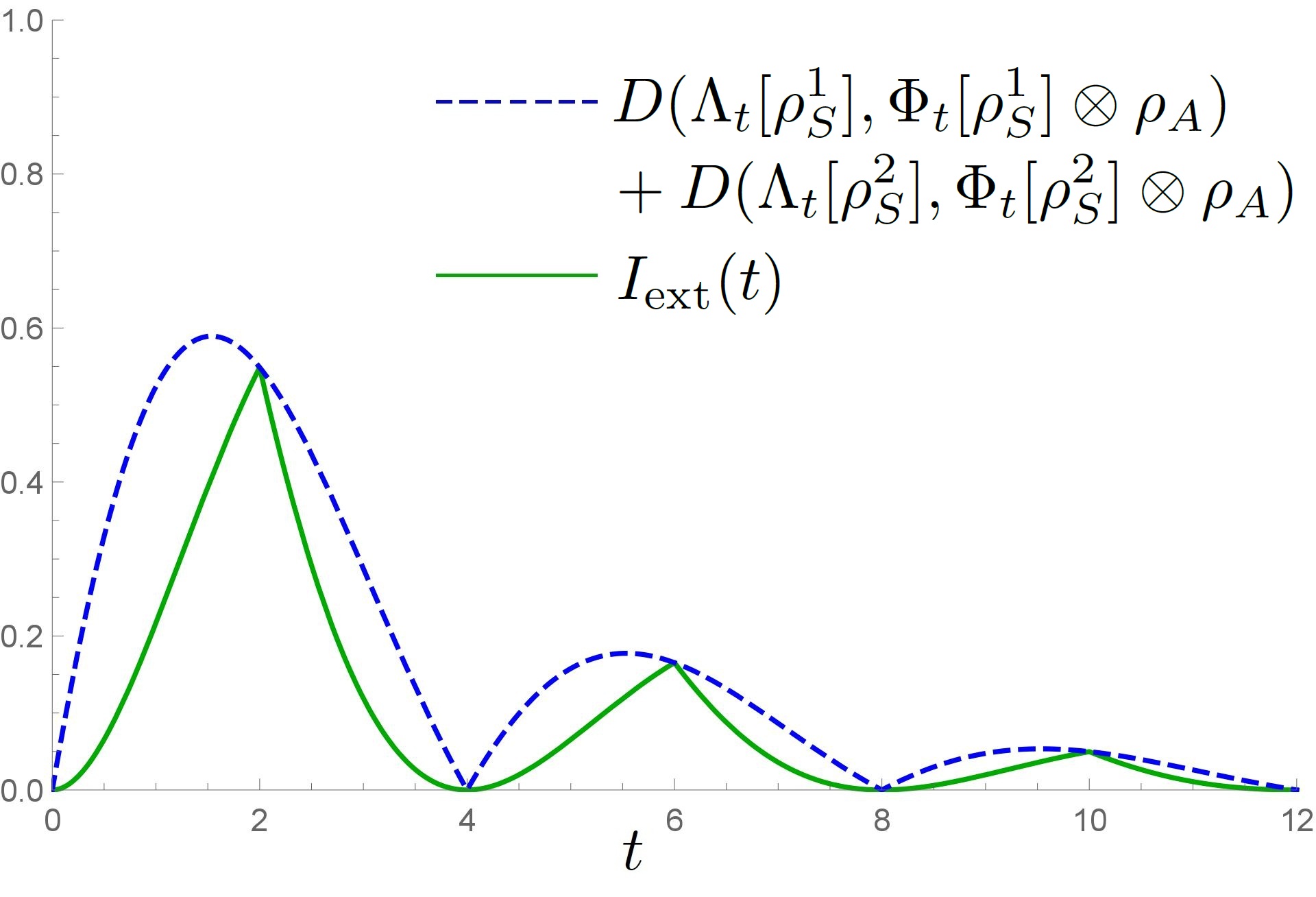}
\caption{Illustration of inequality \eqref{ineq-I_ext-2}. Note that the bound for the external 
information (\ref{eq:extInfo}) is tight, as the equality sign in \eqref{ineq-I_ext-2} holds
whenever $\cos(\pi t/2)=\pm 1$.}
\label{fig:EBdashed}
\end{figure}

Finally, we consider the external information
\begin{eqnarray}
 I_{\textrm{ext}} (t) &=& I_{\textrm{tot}} (t) - I_{\textrm{int}} (t) \nonumber \\
 &=& e^{-t/3} \left[ 1 - |\cos(\pi t/4)| \right]. \label{eq:extInfo}
\end{eqnarray}
This quantity satisfies inequality (22) of the main text:
\begin{equation} \label{ineq-I_ext-2}
 I_{\mathrm{ext}}(t) \leq
 D(\Lambda_t[\rho_S^1] , \Phi_t [\rho_S^1] \otimes \rho_A )
 + D(\Lambda_t[\rho_S^2] , \Phi_t [\rho_S^2] \otimes \rho_A ).
\end{equation}
In the present case the right-hand side of this inequality is found to be 
(see Sec.~\ref{subseq-proof})
\begin{eqnarray}
 && D(\Lambda_t[\rho_S^1] , \Phi_t [\rho_S^1] \otimes \rho_A )
 + D(\Lambda_t[\rho_S^2] , \Phi_t [\rho_S^2] \otimes \rho_A )  \nonumber \\
 && = e^{-t/3} |\sin(\pi t/4)|.
\label{eq:boundExtInfoEsempio}
\end{eqnarray}
Inequality \eqref{ineq-I_ext-2} is illustrated in Fig.~\ref{fig:EBdashed}.

\subsection{Proof of Eq.~(\ref{eq:boundExtInfoEsempio})} \label{subseq-proof}

{\textit{Lemma.}} Suppose we have a bipartite Hilbert space $\mathcal{H}_{S} \otimes
\mathcal{H}_{A}$, a probability distribution $\{ q_{i} \}$, a collection of
statistical operators $\{ \rho^{i}_{S} \}$ on $\mathcal{H}_{S}$, and a
collection of orthogonal projections $\{ \Pi_{i} \}$ on $\mathcal{H}_{A}$,
where $i=1,2,\ldots,n$. Then, for composite states of the form
\begin{equation}
  \rho_{SA} = \sum_{i} q_{i} \rho^{i}_{S} \otimes \Pi_{i}
\end{equation}
the trace distance between the state and the product of its marginals obeys
the bound
\begin{equation}
  D ( \rho_{SA} , \rho_{S} \otimes \rho_{A} )  \leq  2 \sum_{i>j} q_{i}
  q_{j} D ( \rho^{i}_{S} , \rho^{j}_{S} ) ,
\end{equation}
which is saturated for the case $n=2$
\begin{equation}
  D ( \rho_{SA} , \rho_{S} \otimes \rho_{A} ) = 2q_{1} q_{2} D (
  \rho^{1}_{S} , \rho^{2}_{S} ).
\end{equation}
{\textit{Proof.}} Orthogonality of the projections leads to the following
identities
\begin{eqnarray}
  D ( \rho_{SA} , \rho_{S} \otimes \rho_{A} ) & = & D ( \sum_{i} q_{i}
  \rho^{i}_{S} \otimes \Pi_{i} , \sum_{j} q_{j} \rho^{j}_{S} \otimes
  \sum_{i} q_{i} \Pi_{i} ) \nonumber\\
  & = & \frac{1}{2} \| \sum_{i} q_{i} \rho^{i}_{S} \otimes \Pi_{i} -
  \sum_{i,j} q_{j} q_{i} \rho^{j}_{S} \otimes \Pi_{i} \| \nonumber\\
  & = & \frac{1}{2} \| \sum_{i} q_{i} \{ ( 1-q_{i}^{} )
  \rho^{i}_{S} - \sum_{j \neq i} q_{j} \rho^{j}_{S} \} \otimes \Pi_{i}
  \| \nonumber\\
  & = & \frac{1}{2} \sum_{i} q_{i} \| ( 1-q_{i}^{} ) \rho^{i}_{S} -
  \sum_{j \neq i} q_{j} \rho^{j}_{S} \| \nonumber\\
  & = & \frac{1}{2} \sum_{i} q_{i} \| \sum_{j \neq i} q_{j} (
  \rho^{i}_{S} - \rho^{j}_{S} ) \| . 
\end{eqnarray}
For $n=2$ a single term remains and we have the identity
\begin{equation} \label{lemma:n=2}
  D ( \rho_{SA} , \rho_{S} \otimes \rho_{A} ) = 2q_{1} q_{2} D (
  \rho^{1}_{S} , \rho^{2}_{S} ) ,
\end{equation}
for the general case the triangular inequality implies
\begin{equation}
  D ( \rho_{SA} , \rho_{S} \otimes \rho_{A} ) \leq 2 \sum_{i>j} q_{i}
  q_{j} D ( \rho^{i}_{S} , \rho^{j}_{S} ),
\end{equation}
which proves the lemma.

Using Eq.~(\ref{lemma:n=2}) for $q_1=q_2=1/2$ we find
\begin{eqnarray}
 && D(\Lambda_t[\rho_S^1] , \Phi_t [\rho_S^1] \otimes \rho_A )
 + D(\Lambda_t[\rho_S^2] , \Phi_t [\rho_S^2] \otimes \rho_A ) \\ 
 && = \frac{1}{2}   D ( \Phi^{(1)}_t [ \rho_S^1 ], \Phi^{(2)}_t [ \rho_S^1 ] )
 + \frac{1}{2} D ( \Phi^{(1)}_t [ \rho_S^2 ], \Phi^{(2)}_t [ \rho_S^2 ] ) . \nonumber
\label{eq:boundExtInfoParz}
\end{eqnarray}
Because of the symmetric time evolution of $ \rho_S^1 $ and $\rho_S^2 $ 
under $ \Phi^{(1)}_t $ and $ \Phi^{(2)}_t $ (see Figs.~\ref{fig:new_tabella} and \ref{fig:elicheFinal}) we have
\begin{equation}
D ( \Phi^{(1)}_t [ \rho_S^1 ], \Phi^{(2)}_t [ \rho_S^1 ] ) 
= D ( \Phi^{(1)}_t [ \rho_S^2 ], \Phi^{(2)}_t [ \rho_S^2 ] )
\end{equation}
and thus 
\begin{eqnarray}
&& D(\Lambda_t[\rho_S^1] , \Phi_t [\rho_S^1] \otimes \rho_A )
 + D(\Lambda_t[\rho_S^2] , \Phi_t [\rho_S^2] \otimes \rho_A )
 \nonumber \\ 
&& = D ( \Phi^{(1)}_t [ \rho_S^1 ], \Phi^{(2)}_t [ \rho_S^1 ] ) . \label{eq:DD}
\end{eqnarray}
The trace distance $D ( \Phi^{(1)}_t [ \rho_S^1 ], \Phi^{(2)}_t [ \rho_S^1 ] )$ is easily
found to be given by the expression
\begin{equation}
 D ( \Phi^{(1)}_t [ \rho_S^1 ], \Phi^{(2)}_t [ \rho_S^1 ] ) 
 = e^{-t/3} |\sin(\pi t/4)|.
\end{equation}
Substituting this into Eq.~\eqref{eq:DD} we obtain 
Eq.~(\ref{eq:boundExtInfoEsempio}).

\bibliography{mixing}

\end{document}